\documentclass[
 reprint,
 amsmath,amssymb,
 aps,
]{revtex4-2}

\usepackage{graphicx}% Include figure files
\usepackage{newtxmath}
\usepackage{bm}% bold math
\usepackage[colorlinks=true, linkcolor=blue,anchorcolor=blue, citecolor=blue,urlcolor=blue]{hyperref}

\usepackage[normalem]{ulem}

\usepackage{xcolor}

\begin{document}
\preprint{APS/123-QED}

\title{Magnetization-dependent and stacking-tunable Edelstein effect in two-dimensional magnet 2H-VTe$_{2}$}
\author{Weiyi Pan$^{1}$}
\email{Weiyi.Pan@physik.uni-regensburg.de}
\author{Jaroslav Fabian$^{1,2}$}

\affiliation{$^{1}$Institute for Theoretical Physics, University of Regensburg, 93040 Regensburg, Germany\\
$^{2}$Halle-Berlin-Regensburg Cluster of Excellence CCE, University of Regensburg, 93040 Regensburg, Germany\\
}

\begin{abstract}
The Edelstein effect in magnetic systems enables magnetization switching via the coupling between current-induced spin accumulation and intrinsic magnetic order, and is therefore highly promising for next-generation spintronic devices. Realizing and manipulating the Edelstein effect in two-dimensional (2D) magnetic systems is particularly desirable for achieving high-efficiency and multifunctional spintronic applications. In this work, based on first-principles calculations and symmetry analysis, we demonstrate that the Edelstein effect can intrinsically arise in the 2D in-plane ferromagnetic semiconductor 2H-VTe$_{2}$, with its behavior strongly dependent on the magnetization orientation. For monolayer 2H-VTe$_{2}$ with $D_{3h}$ crystal symmetry, under an applied current along the +$x$ direction, only the time-reversal-even $z$ component and the time-reversal-odd $y$($x$) component of the spin accumulation are allowed when the magnetization is aligned along +$x$ (+$y$). For ferromagnetic bilayer 2H-VTe$_{2}$ in AB or BA stacking, where the crystal symmetry is reduced to $C_{3v}$, additional spin components emerge with the presence of in-plane magnetization. Specifically, for magnetization along +$x$ (+$y$), besides $dS_z^{\mathrm{even}}$ and $dS_y^{\mathrm{odd}}$($dS_z^{\mathrm{even}}$ and $dS_x^{\mathrm{odd}}$), extra components such as $dS_y^{\mathrm{even}}$and $dS_z^{\mathrm{odd}}$($dS_y^{\mathrm{even}}$) become allowed. Notably, these additional components can be reversibly switched by changing the stacking configuration from AB to BA via interlayer sliding, which is equivalent to applying a $M_{z}T$ operation. The parity dependence of Edelstein coefficients on the in-plane magnetization is also evaluated. Our results not only deepen the understanding of current-induced spin accumulation in 2D magnetic systems from both symmetry and first-principles perspectives, but also identify 2H-MX$_{2}$ materials as a promising platform for realizing intrinsic and tunable Edelstein effects in high-efficiency spin–orbit torque devices.

\end{abstract}
\maketitle
\section{Introduction}

In non-centrosymmetric systems with strong spin–orbit coupling (SOC), an electric current can induce a nonequilibrium spin accumulation, known as the Edelstein effect\cite{EDE,Johansson_2024}. In magnetic systems, such current-induced spin accumulation can exert spin–orbit torques on the intrinsic magnetic moments via exchange coupling\cite{transfer}, thereby enabling electrical manipulation of the magnetization orientation\cite{RevModPhys.91.035004,RevModPhys.96.015005}. This mechanism further underpins various spintronic applications, such as magnetoresistive random-access memory (MRAM) based on magnetization switching and racetrack memory based on domain-wall motion\cite{9427163,berger1984exchange,SONG2021100761}. A straightforward strategy to realize the Edelstein effect in magnetic systems is to construct heterostructures composed of ferromagnetic (FM) metals and nonmagnetic materials with strong SOC. In such systems, inversion symmetry is broken at the interface, giving rise to a nontrivial Edelstein effect\cite{E1,E2,Kubo1,HM1,HM2,HM3,HM4}. However, in heterostructures with finite thickness,
the interaction between current-induced spin accumulation and localized magnetic moments is typically confined near the interface\cite{E2,E1,E3}, rather than extending throughout the entire system, thereby reducing the efficiency of the resulting spin–orbit torque. 
Moreover, the finite thickness of such heterostructures not only complicates fabrication but also limits device integration density. Therefore, it is highly desirable to realize the Edelstein effect in intrinsically atomic-thin two-dimensional (2D) magnetic systems with broken inversion symmetry, where the interaction between current-induced spin accumulation and magnetic moment can extend across the entire material. Such systems can not only enhance the efficiency of current-driven magnetic control, but also improve the device integration density.

Beyond realizing the intrinsic Edelstein effect in non-centrosymmetric 2D magnets, another key objective is its controllable manipulation, which is essential for tunable spintronic devices. Since the Edelstein effect is fundamentally governed by crystal symmetry, the individual components of current-induced spin accumulation can in principle be selectively activated, suppressed, or even reversed through deliberate symmetry engineering\cite{kurebayashi2022magnetism,REE3,REE4,REE5}. This tunability enables possible diversity in spin–orbit torque behaviors and thus provides a versatile platform for multifunctional spintronic applications. In 2D magnetic systems, crystal symmetry can be effectively engineered via bilayer stacking\cite{chen2024twist,yang2024macroscopic,PhysRevLett.133.206702,dmzg-ck2t,PhysRevLett.130.146801,PhysRevLett.133.166701}, which often reduces the symmetry of the whole system compared to the monolayer case and thereby introduces additional components of the Edelstein effect. Moreover, electrical switching between distinct bilayer stacking configurations can be achieved through sliding ferroelectricity\cite{PhysRevLett.130.146801,wu2021sliding}, which is in principle able to switch the sign of Edelstein effect and thus offering a promising route toward nonvolatile spintronic devices\cite{FE4}. However, such stacking-dependent Edelstein effect in 2D magnets remains largely unexplored, thereby motivating further investigation. 

2D FM transition-metal compounds MX$_{2}$ (M = magnetic elements, X = nonmagnetic elements) in the 2H phase (e.g., VSe$_{2}$\cite{MX2_1,MX2_2,MX2_3,MX2_4,MX2_5,MX2_6}, VTe$_{2}$\cite{VStrain,MX2_4,MX2_6}, FeCl$_{2}$\cite{MX2_7,MX2_8,MX2_10}, and FeBr$_{2}$\cite{MX2_8,MX2_9}), which have attracted considerable attention due to their rich magnetism-related properties, provide a simple yet powerful platform for addressing the aforementioned two objectives. In particular, their crystal symmetry belongs to the $D_{3h}$ point group, which intrinsically lacks inversion symmetry and further allows the emergence of the current-induced spin accumulation in the presence of magnetism and SOC\cite{FGT,FGT2,FGT3}. Moreover, the electronic structure of 2H-MX$_{2}$ is strongly coupled to magnetism, giving rise to a variety of magnetization-dependent physical properties, such as ferrovalley effects\cite{MX2_2,MX2_8}, multiferroicity\cite{MX2_4,PhysRevMaterials.8.104403}, and magnetic topological states\cite{MX2_10,MX2_7}. Accordingly, the Edelstein effect in these systems is expected to be strongly dependent on the magnetization orientation, likely governed by the corresponding magnetic group symmetry, and thus warrants further systematic investigation. Furthermore, when 2H-MX$_{2}$ layers are stacked into bilayers with specific polar stackings, ferroelectric polarization can emerge\cite{VStrain,MX2_11,MX2_12}, enabling electrical switching between these two distinct stacking configurations. Such switching may lead to a reversal of the Edelstein effect, which calls for a detailed symmetry-based analysis.

In the present work, we focus on 2H-VTe$_2$. To date, two distinct phases of VTe$_2$ have been reported or proposed. One is the experimentally accessible metallic 1T phase, which exhibits, in few-layer samples, charge-density-wave distortions at low temperatures\cite{1T-VTe2,1T-Vte22,1T-Vte23}. The other is the theoretically predicted \cite{MX2_13,WANG2022152520} 2H phase, which represents an exciting target for experimental realization owing to the
variety of intriguing properties expected for it \cite{TOP,WANG2022152520,WANG2021148098,MX2_4,VStrain}. In particular, monolayer VTe$_{2}$ is predicted to be a magnetic semiconductor which can host valley polarization and higher-order topological phase \cite{TOP,WANG2022152520,WANG2021148098}. Magnetism-induced ferroelectricity can emerge upon changing the magnetization direction, making 2H-VTe$_2$ a rare example of a type-II multiferroic system\cite{MX2_4}. Moreover, sliding ferroelectricity has been predicted in bilayer 2H-VTe$_2$, enabling the valley polarization to be controlled by reversing the electric polarization through stacking changes\cite{VStrain}. It is therefore of interest to theoretically explore the magnetic- and stacking-dependent Edelstein effect in 2H-VTe$_2$, which may help to reveal further exotic phenomena in this system and provide guidance for future experimental studies. 

 \begin{figure*}[ht]
\includegraphics[scale = 0.33 ]{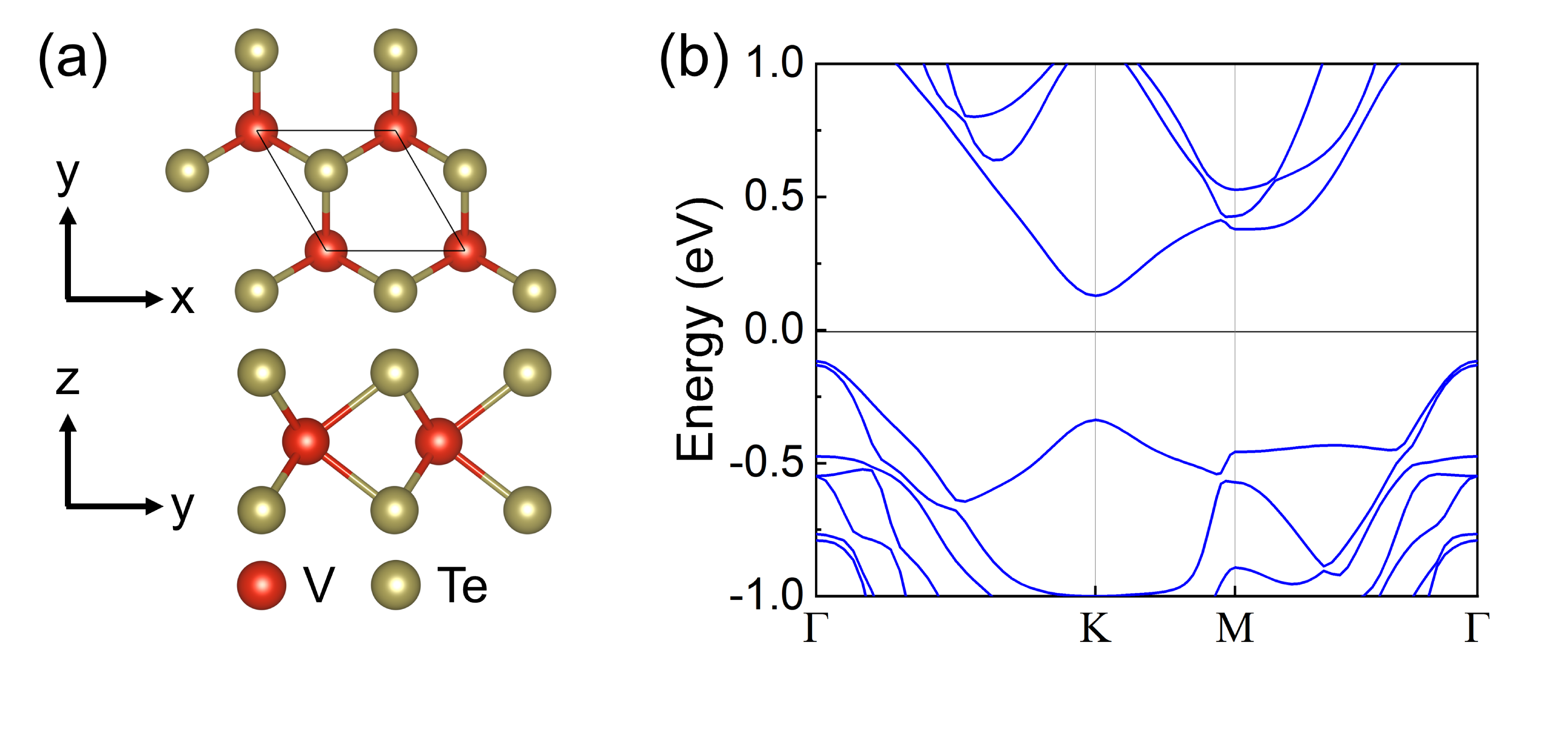}
\caption{\label{1} (a) The atomic structure of monolayer 2H-VTe$_{2}$. (b) The band structure of monolayer 2H-VTe$_{2}$.  }
\end{figure*}

In this work, using first-principles calculations combined with symmetry analysis, we systematically investigate the Edelstein effect and its relationship with magnetic symmetry in 2H-VTe$_{2}$, which is known to be a ferromagnetic semiconductor in the family of 2H-MX$_{2}$\cite{MX2_6,MX2_4}. For monolayer 2H-VTe$_{2}$, we find that in-plane magnetization lowers the symmetry, thereby giving rise to a nontrivial Edelstein effect. Notably, this effect strongly depends on the magnetization orientation. For example, when the magnetization is aligned along the +$x$ (+$y$) direction, only the time-reversal-even $z$ component and the time-reversal-odd $y$ component (the time-reversal-even $z$ component and the time-reversal-odd $x$ component) of the spin accumulation are allowed by magnetic symmetry under an electric field applied along the +$x$ direction. For ferromagnetic bilayer 2H-VTe$_{2}$ in AB and BA stacking configurations, additional components of current-induced spin accumulation emerge due to the further reduction of crystal symmetry, some of which can be reversed upon switching between AB and BA stacking. Moreover, rotating the in-plane magnetization leads to symmetry-constrained angular dependence of the spin accumulation in both monolayer and bilayer systems. Our results not only provide symmetry-based insights into charge–spin conversion in 2D magnetic systems, but also expand the range of candidate materials for high-efficiency spin–orbit torque devices.

\section{Methods}
First-principles calculations are performed using the QUANTUM ESPRESSO package\cite{QE}. The Perdew–Burke–Ernzerhof (PBE) exchange–correlation functional and projector augmented-wave (PAW) potentials are employed\cite{PAW,PBE}. Both structural relaxations and self-consistent calculations are carried out using an $18 \times 18 \times 1$ k-point mesh, with plane-wave energy cutoffs of 60 Ry for the wavefunctions and 700 Ry for the charge density. Structural relaxations are performed using scalar-relativistic pseudopotentials and the quasi-Newton algorithm until all force components are smaller than $2\times{10}^{-4}\textup{Ry}/a_0$, where $a_0$ denotes the Bohr radius.  Grimme D-2 van der Waals corrections is adopted during the relaxation\cite{DFT-D3,DFT-D2}. Subsequently, self-consistent calculations are conducted with fully relativistic pseudopotentials to include SOC. A vacuum region of at least 20 \AA is introduced to eliminate spurious interactions between periodic images in the slab geometry.  To account for the on-site Coulomb interaction of the V-3$d$ orbitals, a Hubbard U parameter of 1 eV is applied, consistent with previous studies\cite{VStrain}. 

After the density functional theory (DFT) calculations, we construct the Wannier Hamiltonian using the Wannier90 package\cite{w901}. In the wannierization procedure, the V-$d$ orbitals and Te-$p$ orbitals are chosen as the projection basis. Based on the resulting Wannier Hamiltonian, the WannierTools package is employed to compute the Fermi surface\cite{wtools}. Furthermore, the Edelstein coefficient is evaluated using the Linres code within the framework of the Kubo formula\cite{linears}. Specifically, we consider the linear response of spin accumulation to an external electric field, expressed as $\delta \mathbf{S}=\chi \mathbf{E}$, where $\delta \mathbf{S}$ denotes the induced spin accumulation, $\mathbf{E}$ is the applied electric field, and $\chi$ is the Edelstein response tensor.
Within the Kubo formalism, $\chi$ can be decomposed into time-reversal-even and time-reversal-odd contributions, respectively\cite{Kubo1}:

\begin{equation}
    \chi^{\mathrm{even}}_{ij} = -\frac{e\hbar}{\pi  N} \sum_{\mathbf{k},m,n} \frac{\Gamma^{2} \textup{Re}(\langle n\mathbf{k} |\hat{S}_{i} | m\mathbf{k} \rangle \langle m\mathbf{k} | \hat{v}_{j} | n\mathbf{k} \rangle  )  }{[(E_{f}-\epsilon_{n\mathbf{k}})^{2}+ \Gamma^{2}][(E_{f}-\epsilon_{m\mathbf{k}})^{2}+ \Gamma^{2}]}
\end{equation}

\begin{equation}
\begin{aligned}
 \mathbf{\chi}_{ij}^{\mathrm{odd}}
&= \frac{2e\hbar}{N}
\sum_{\mathbf{k},\,n\neq m}^{\substack{n \   \mathrm{occ.} \\ m \ \mathrm{unocc.}}}
\operatorname{Im}
\Big[
\langle nk|\hat{S}_{i}|mk\rangle \notag  
\langle mk|\hat{v}_{j}|nk\rangle
\Big] \notag \\
&\quad \times
\frac{\Gamma^{2}-\left(\epsilon_{nk}-\epsilon_{mk}\right)^{2}}
{\left[\left(\epsilon_{nk}-\epsilon_{mk}\right)^{2}+\Gamma^{2}\right]^{2}}.
\end{aligned}
\end{equation}

Here, $e$ is the elementary charge; $n,m$ denote band indices; $\mathbf{k}$ is the Bloch vector; $N$ is the total number of k points used to sample the Brillouin zone; $E_F$ is the Fermi energy; $\mathbf{V}$ is the velocity operator; $\varepsilon_{n\mathbf{k}}$ is the eigenvalue; and $\Gamma$ is the disorder parameter, which is related to the relaxation time $\tau$ through: $\tau = \hbar/2\Gamma$. It should be noted that the time-reversal-even component of the Edelstein effect originates solely from electronic states at the Fermi surface and is also termed as “intraband” contribution. This can be understood more clearly in the limit of a small broadening parameter $\Gamma$, where the corresponding expression can be approximated as\cite{Kubo1}:

\begin{equation}
    \chi_{ij}^{\textup{even}} = -\frac{e\hbar}{2 \Gamma  N} \sum_{\textbf{k},n}     \delta(\epsilon_{nk}- E_{F}) 
 \langle n\textbf{k} | \hat{S}_{i} | n\textbf{k} \rangle \langle n\textbf{k} |\hat{v}_{j} | n\textbf{k} \rangle  
\end{equation}

where the $\delta$-function explicitly indicates that only states at the Fermi level contribute to the Edelstein effect. 
Meanwhile, the time-reversal-odd part of Edelstein coefficient, which is also termed as “interband” contribution, goes to the $\Gamma$-independent intrinsic formula when $\Gamma$ is very small\cite{Kubo1}:
\begin{equation}
\chi_{ij}^{\mathrm{odd}}
=
-\frac{2e\hbar}{N}
\sum_{\mathbf{k},\, n\neq m}^{\substack{n\in \mathrm{occ.} \\ m\in \mathrm{unocc.}}}
\frac{
\operatorname{Im}
\!\left[
\langle nk|\hat{S}_{i}|mk\rangle
\langle mk|\hat{v}_{j}|nk\rangle
\right]
}{
(\epsilon_{nk}-\epsilon_{mk})^{2}
}.
\end{equation}

In this work, for a clear illustration of the current-induced spin accumulation, we restrict the direction of electric field along +$x$ direction, i.e., $j$ = $x$. In this case, the calculated Edelstein coefficient components $\chi_{ix}$ is directly associated with the current-induced spin accumulation components in a unit cell along $i$ ($i = x,y,z$)  direction, $\delta S_{i}$. During the calculation, we set $\Gamma$=10 meV, which roughly corresponds to a typical sub-100 fs scattering rate (broadening) while remaining much smaller than interband separations. $600\times600\times1$ kpoints are adopted to arrive at the converged EE coefficient.

\section{Edelstein effect in monolayer 2H-${\mathrm{VTe}_{2}}$}

 \begin{figure*}[ht]
\includegraphics[scale = 0.46 ]{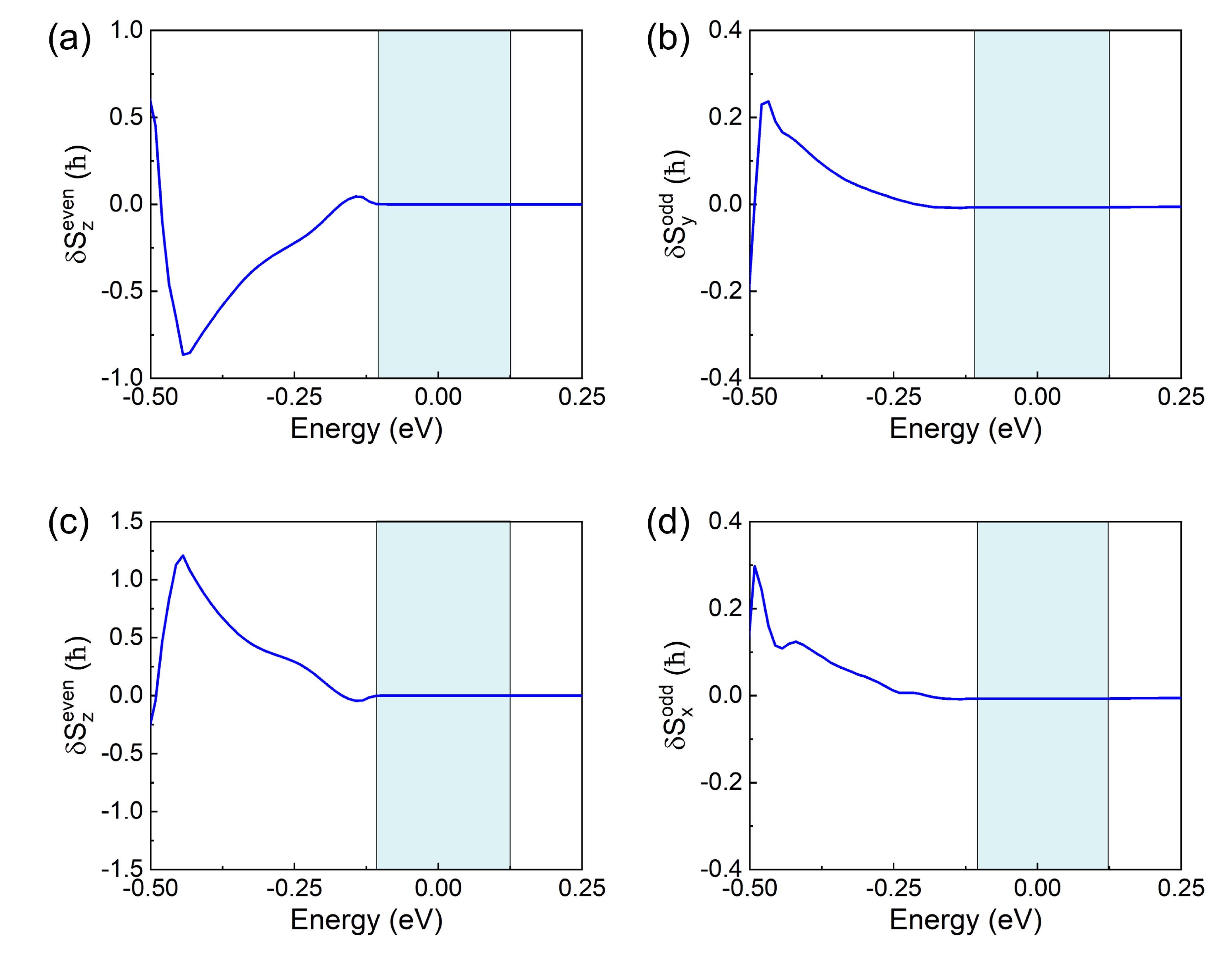}
\caption{\label{2} Calculated symmetry-allowed current-induced spin accumulations in a unit cell as functions of Fermi energy in monolayer 2H-VTe$_{2}$ under a reference  electric field 1 V/\AA along +$x$ direction. (a) and (b) show the time-reversal-even ($\delta S_z^{\mathrm{even}}$) and time-reversal-odd ($\delta S_y^{\mathrm{odd}}$) components of spin accumulation when magnetization is along +$x$ direction. (c) and (d) show the time-reversal-even ($\delta S_z^{\mathrm{even}}$) and time-reversal-odd ($\delta S_x^{\mathrm{odd}}$) components of spin accumulation when magnetization is along +$y$ direction.  The dashed blue region indicates the bandgap. Note that the magnitude of the electric field (1 V/\AA) is difficult to realize experimentally and is used only as a reference value.}  
\end{figure*}

 \begin{figure*}[ht]
\includegraphics[scale = 0.42 ]{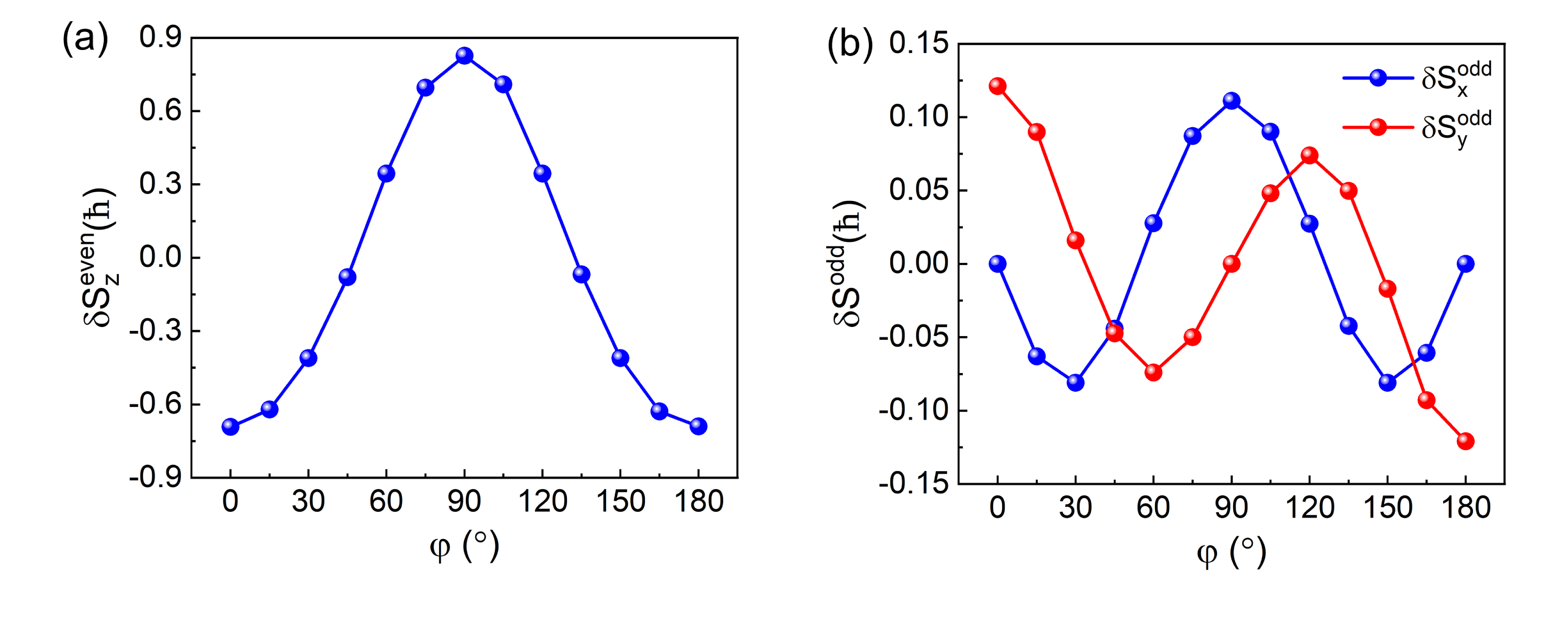}
\caption{\label{3} Angular-dependent spin accumulation in a unit cell of monolayer 2H-VTe$_{2}$, which is denoted as a function of $\phi$ (angle between magnetization direction and +$x$ direction). (a) denotes the symmetry-allowed time-reversal-even components (only $\delta S_z^{\mathrm{even}}$ here) of spin accumulation in a unit cell as function of $\phi$, while (b) denotes the symmetry-allowed time-reversal-odd components of spin accumulation in a unit cell as function of $\phi$. Note that the magnetization is rotating in plane, which means that $\phi = 0$ and $\phi = 90^\circ$ denotes +$x$ and +$y$ magnetization, respectively. The spin accumulation is calculated at $E_{F}$ = -0.4 eV under a fixed electric field of 1 V/\AA along +$x$ direction. Note that this magnitude of electric field is difficult to realize experimentally and is used only as a reference value.}  
\end{figure*}

The crystal structure of monolayer 2H-VTe$_{2}$ is shown in Fig. \ref{1}(a), which crystallizes in space group No. 187 ($D_{3h}$ point group). Previous studies have predicted that monolayer 2H-VTe$_{2}$ is a stable ferromagnetic (FM) semiconductor\cite{MX2_4,MX2_6,MX2_13}. As illustrated, the V atoms form a triangular lattice, with a nearest-neighbor V–V distance of 3.61 \AA after full structural relaxation. Each V atom is coordinated by six Te atoms, forming a local trigonal prismatic environment. This trigonal crystal field splits the V 3$d$ orbitals into three groups: the $d_{z^2}$ orbital (denoted as $A_1$), the degenerate $d_{xy}$ and $d_{x^2-y^2}$ orbitals (denoted as $E_1$), and the $d_{xz}$ and $d_{yz}$ orbitals (denoted as $E_2$). The calculated magnetic moment is 1 $\mu_B$ per unit cell, consistent with the 3$d^1$ electronic configuration of each V$^{4+}$ ion. Furthermore, the magnetic anisotropy energy (MAE) is evaluated as $E_{x}-E_{z} = E_{y}-E_{z}$ = -2.07 meV per V atom, indicating that the in-plane magnetization is energetically favored over the out-of-plane orientation. This result is consistent with previous reports\cite{MX2_4,MX2_6}. Therefore, in the following, we focus on the in-plane magnetization states of monolayer 2H-VTe$_{2}$.The electronic band structure of monolayer 2H-VTe$_{2}$ is presented in Fig.  \ref{1}(b). The valence band maximum (VBM) is located at the $\Gamma$ point, while the conduction band minimum (CBM) resides at the $K$ point, resulting in an indirect band gap of approximately 0.23 eV.

Now let us qualitatively explore the Edelstein effect in monolayer 2H-VTe$_{2}$ from the viewpoint of symmetry. 
Considering only the $D_{3h}$ crystal symmetry of monolayer 2H-VTe$_{2}$ without the inclusion of magnetism, the Edelstein effect is forbidden. However, the introduction of in-plane magnetization breaks the $D_{3h}$ symmetry, thereby allows a finite current-induced spin accumulation.
For instance, when the magnetization is aligned along the +$x$ direction, the magnetic point group reduces to $m^\prime m2^\prime$, with symmetry generators $\{IM_{x},C_{2y}T, M_{z}T\}$. Here $I$ denotes spatial inversion operation, $T$ denotes time reversal operation, $M_{\alpha}$ ($\alpha = x,y,z$) denotes the mirror operation with the normal of mirror plane along $\alpha$ direction, while $C_{2\alpha}$ ($\alpha = x,y,z$) denotes the two-fold rotation operation with respect to $\alpha$ axis. Analysis of these symmetry operations shows that the $M_x$ operator reverses the sign of $\chi_{xx}^{\mathrm{even}}$ and $\chi_{xx}^{\mathrm{odd}}$, while $C_{2y}T$ reverses $\chi_{yx}^{\mathrm{even}}$, and $M_zT$ reverses $\chi_{zx}^{\mathrm{odd}}$. Consequently, $\chi_{xx}^{\mathrm{even}}$, $\chi_{xx}^{\mathrm{odd}}$, $\chi_{yx}^{\mathrm{even}}$, and $\chi_{zx}^{\mathrm{odd}}$ are forbidden under the $m^\prime m2^\prime$ symmetry. In contrast, $\chi_{zx}^{\mathrm{even}}$and $\chi_{yx}^{\mathrm{odd}}$remain invariant under all symmetry operations and are therefore allowed. As a result, when the magnetization lies along the +$x$ direction, only $\delta S_z^{\mathrm{even}}$and $\delta S_y^{\mathrm{odd}}$ are induced under an electric field applied along the +$x$ direction.
Similarly, when the magnetization is aligned along the +$y$ direction, the magnetic point group becomes $m^\prime m^\prime2$, with symmetry operators $\{I,C_{2y},M_xT, M_zT\}$. Symmetry analysis yields the following constraints: (i) $M_xT$ forbids $\chi_{xx}^{\mathrm{even}}$and $\chi_{yx}^{\mathrm{odd}}$; (ii) $M_zT$ forbids $\chi_{zx}^{\mathrm{odd}}$; (iii) $C_{2y}$forbids $\chi_{yx}^{\mathrm{even}}$. Meanwhile, $\chi_{zx}^{\mathrm{even}}$ and $\chi_{xx}^{\mathrm{odd}}$ remain invariant under all symmetry operations and are thus allowed. Therefore, for magnetization along the +$y$ direction, only $\delta S_z^{\mathrm{even}}$ and $\delta S_x^{\mathrm{odd}}$ are induced under an electric field applied along the +$x$ direction.
Notably, the above symmetry-based analysis is general and can be extended to other 2H-MX$_{2}$ systems with in-plane magnetization.

To gain quantitative insights into the Edelstein effect in monolayer 2H-VTe$_{2}$, we calculate the symmetry-allowed components of current-induced spin accumulation per unit electric field (V/\AA) as a function of the Fermi energy, as shown in Fig. \ref{2}. We first consider the case where the magnetization is aligned along the +$x$ direction, for which only $\delta S_z^{\mathrm{even}}$ and $\delta S_y^{\mathrm{odd}}$ are allowed [see Fig. \ref{2}(a) and \ref{2}(b)]. For $\delta S_z^{\mathrm{even}}$, its magnitude increases as the Fermi energy shifts downward toward the valence band, reaching a maximum value of about -0.86 $\hbar$ at $E_F= -0.44$ eV (approximately 0.3 eV below the valence band maximum), which is microscopically dominated by the competing contribution from the states near $\Gamma$ point (see Fig.\ref{7} in Appendix B). Upon further decreasing the Fermi energy, $\delta S_z^{\mathrm{even}}$decreases in magnitude and changes sign around $E_F$=-0.5eV. The behavior of $\delta S_y^{\mathrm{odd}}$ is qualitatively similar, exhibiting an initial increase followed by a decrease and sign reversal as the Fermi level shifts downward, and the local maxima is dominated by the states near the $\Gamma$ point (see Fig. \ref{7} in Appendix B). However, two key differences can be identified: (i) $\delta S_z^{\mathrm{even}}$ and $\delta S_y^{\mathrm{odd}}$ have opposite signs within the valence band region; (ii) the maximum magnitude of $\delta S_y^{\mathrm{odd}}$(about 0.23 $\hbar$) is significantly smaller than that of $\delta S_z^{\mathrm{even}}$ (about 0.86 $\hbar$). Similarly, when the magnetization is aligned along the +$y$ direction, the symmetry-allowed components $\delta S_z^{\mathrm{even}}$and $\delta S_x^{\mathrm{odd}}$are shown in Figs. \ref{2}(c) and \ref{2}(d). As the Fermi energy shifts downward, both components remain positive and increase in magnitude, reaching their respective maxima near $E_F=-0.4$ eV, with values of approximately 1.2 $\hbar$ for $\delta S_z^{\mathrm{even}}$ and about 0.12 $\hbar$ for $\delta S_x^{\mathrm{odd}}$. Notably, for magnetization oriented along either +$x$ or +$y$, both the time-reversal-odd and time-reversal-even spin accumulations are nearly zero in the low-energy conduction-band region. Microscopically, this region is dominated by electronic states near the $K$ points and their symmetry-related partners. The time-reversal-even contributions from these states cancel each other, while the time-reversal-odd spin accumulation remains negligible possibly due to the large interband separation at the $K$ point suppresses the corresponding interband contributions to the spin accumulations (see Fig.\ref{8} in Appendix B).

 \begin{figure*}[ht]
\includegraphics[scale = 0.42 ]{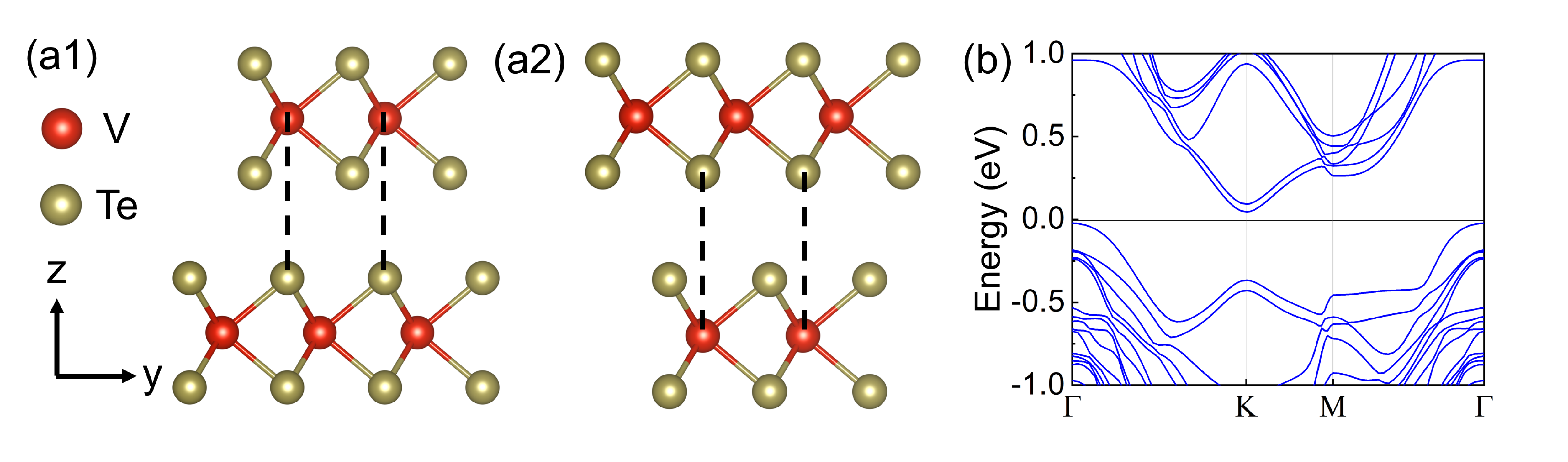}
\caption{\label{4} The atomic structure of bilayer 2H-VTe$_{2}$ in (a) AB and (b) BA stacking, respectively. (c) The band structure of AB-stacked bilayer 2H-VTe$_{2}$, which possesses the same qualitative feature (i.e. CBM at $\Gamma$ and VBM at $k$) with BA-stacked system. }
\end{figure*}

 \begin{figure*}[ht]
\includegraphics[scale = 0.46 ]{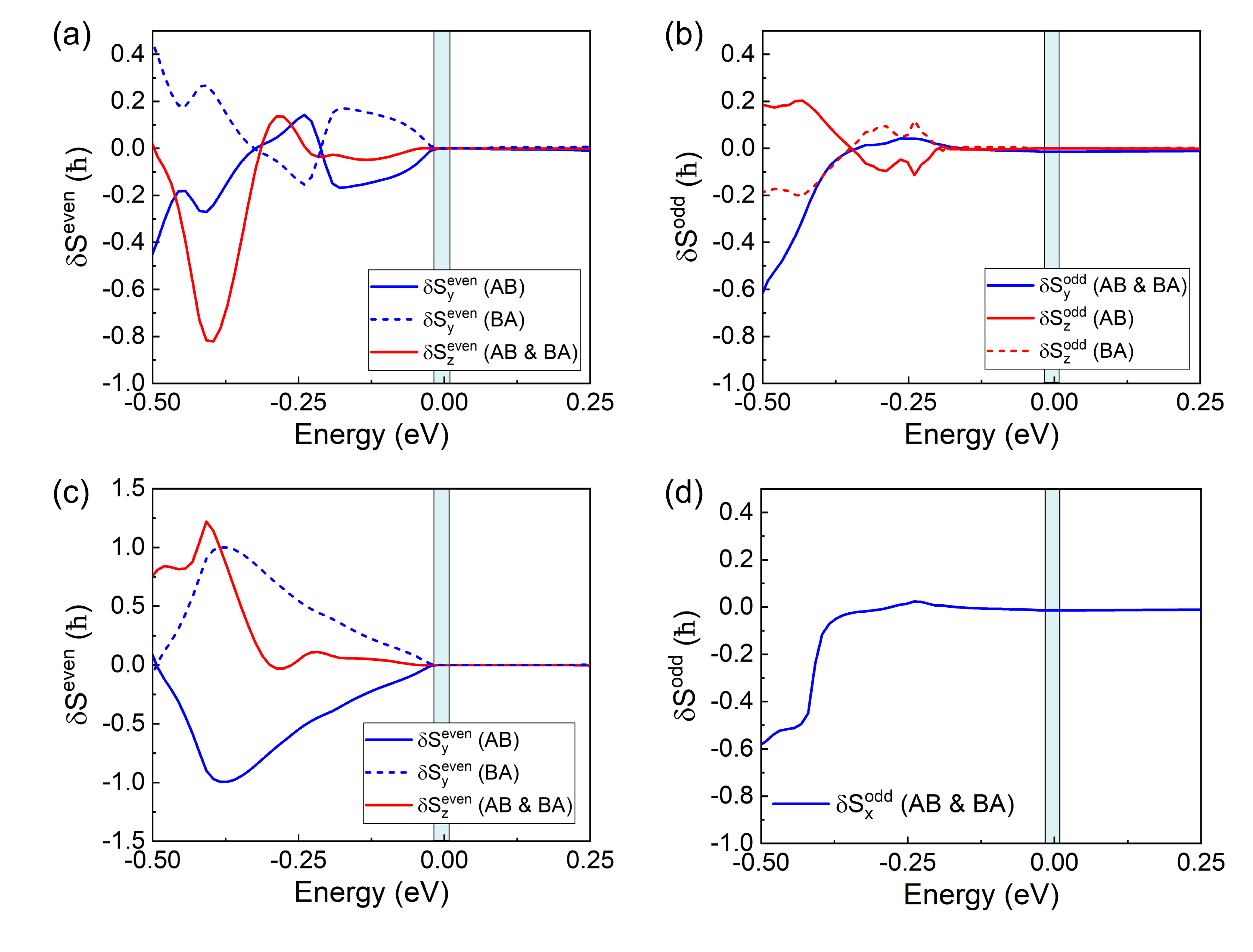}
\caption{\label{5} Calculated symmetry-allowed current-induced spin accumulations in a unit cell as functions of Fermi energy in bilayer 2H-VTe$_{2}$ with distinct stackings under a reference electric field  of 1 V/\AA   along +$x$ direction. (a) and (b) show the time-reversal-even and time-reversal-odd components of spin accumulation when magnetization is along +$x$ direction. (c) and (d) show the time-reversal-even and time-reversal-odd components of spin accumulation when magnetization is along +$y$ direction. AB (BA) denotes the spin accumulation components calculated for AB (BA) stacking in the bilayer system, while AB \& BA means both stackings yield the identical numerical result due to the constraint of symmetry. The dashed blue region indicates the bandgap. Note that the magnitude of the electric field (1 V/\AA) is difficult to realize experimentally and is used only as a reference value.  }
\end{figure*}

Based on the above analysis, it is evident that the current-induced nonequilibrium spin accumulation in monolayer 2H-VTe$_{2}$ strongly depends on the magnetization orientation, warranting further investigation. To this end, we evaluate the time-reversal-even and -odd components of the induced spin at $E_F$=-0.4 eV as a function of the in-plane magnetization angle $\phi$, as shown in Figs. \ref{3}(a) and \ref{3}(b). As $\phi$ varies, only the $z$ component of spin accumulation appears in the time-reversal-even contribution, whereas both $x$ and $y$ components are present in the time-reversal-odd part. When $\phi$ increases from 0 to $180^\circ$, $\delta S_z^{\mathrm{even}}$first increases, reaching a maximum of approximately 0.9 $\hbar$ at $\phi={90}^\circ$, and then decreases symmetrically with respect to $\phi={90}^\circ$. Quatitatively, the $\delta S_z^{\mathrm{even}}$ is roughly proportional to cos2$\phi$, which can be understood from the second-order dependence on the magnetization direction of current-induced spin accumulation (see Appendix C). Similarly, $\delta S_x^{\mathrm{odd}}$ initially decreases from zero, then increases to a local maximum of about $0.12 \hbar$ at $\phi={90}^\circ$, and subsequently exhibits symmetric behavior about $\phi={90}^\circ$. In contrast, $\delta S_y^{\mathrm{odd}}$decreases from approximately $0.12 \hbar$ to a local minimum of about $-0.08 \hbar$, and then increases to zero at $\phi={90}^\circ$, displaying an antisymmetric dependence with respect to $\phi={90}^\circ$. The calculated spin responses, with magnitudes on the order of $0.1 \sim 1 \hbar$ per V/\AA, are comparable to those reported in p-wave magnetic systems such as LuFeO$_{3}$ \cite{Manchon} and CeNiOAs\cite{Libor}. By normalizing with the calculated conductivity and assuming an effective thickness of 6 \AA for the monolayer, we estimate that a current density of $10^7$ A/cm$^2$ induces magnetic moments on the order of ${10}^{-7}$ to ${10}^{-6} \hbar$ per unit cell (see Appendix A). This magnitude is comparable to that of nonmagnetic chiral and ferroelectric systems (e.g., CsGeBr$_{3}$\cite{FE6}, WTe$_{2}$\cite{FE4}, and NbSi$_{2}$\cite{REE17}), and exceeds that predicted for antiferromagnetic Mn$_{2}$Au\cite{FGT3}. To understand the parity dependence of the current-induced spin components on the magnetization angle, we note that switching the magnetization angle between $\phi$ to ${180}^\circ-\phi$ is equivalent to applying the $M_xT$ operation. Under this symmetry, $\chi_{zx}^{\mathrm{even}}$ and $\chi_{xx}^{\mathrm{odd}}$ remain invariant, whereas $\chi_{yx}^{\mathrm{odd}}$ changes sign. Consequently, for an electric field applied along the +$x$ direction, $\delta S_z^{\mathrm{even}}$ and $\delta S_x^{\mathrm{odd}}$, which are proportional to $\chi_{zx}^{\mathrm{even}}$ and $\chi_{xx}^{\mathrm{odd}}$, respectively, are even functions with respect to $\phi-{90}^\circ$, while $\delta S_y^{\mathrm{odd}}$, proportional to $\chi_{yx}^{\mathrm{odd}}$, is an odd function. This analysis establishes a direct connection between the angular dependence of the Edelstein effect and the underlying symmetries, and is expected to be generally applicable to other magnetic systems with similar crystal symmetry. 

 \begin{figure*}[ht]
\includegraphics[scale = 0.42 ]{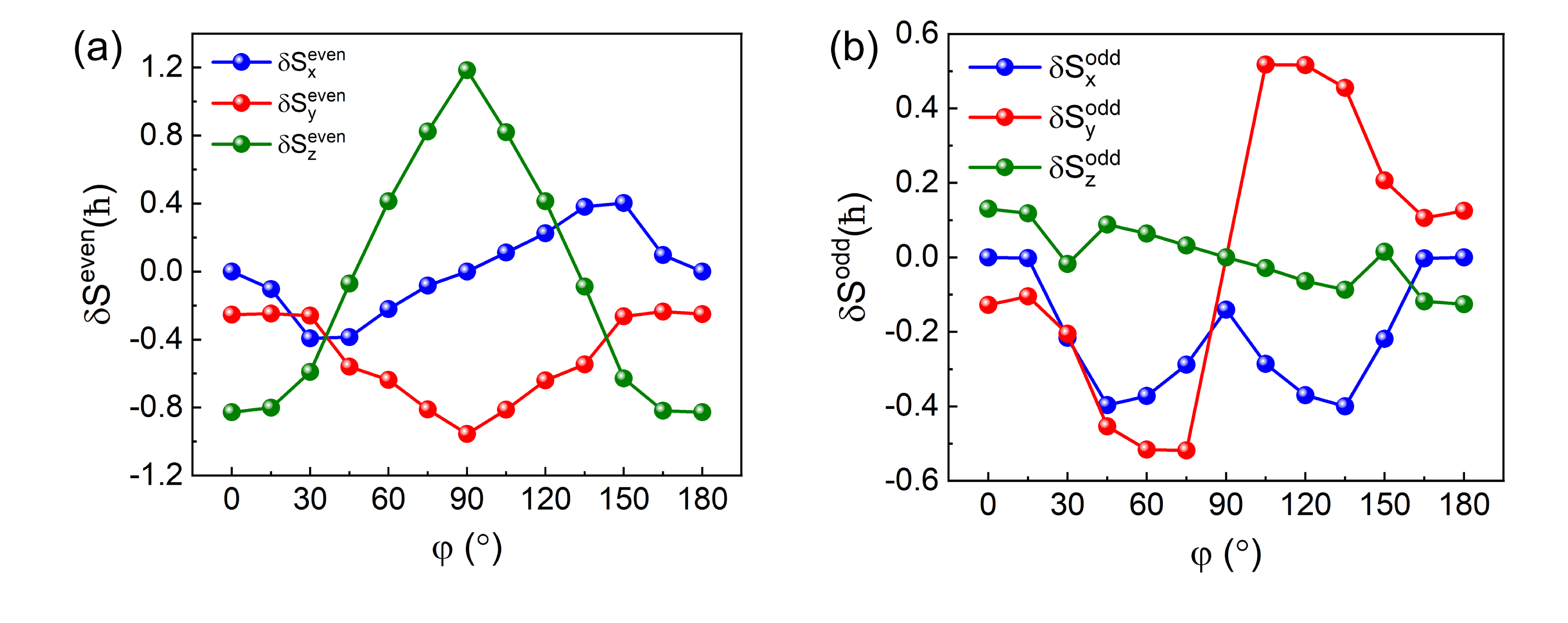}
\caption{\label{6} Angular-dependent spin accumulation in a unit cell of AB-stacked bilayer 2H-VTe$_{2}$, which is denoted as a function of $\phi$ (angle between magnetization direction and +$x$ direction). (a) denotes the symmetry-allowed time-reversal-even components (only $\delta S_z^{\mathrm{even}}$ here) of spin accumulation in a unit cell as function of $\phi$, while (b) denotes the symmetry-allowed time-reversal-odd components of spin accumulation in a unit cell as function of $\phi$. Note that the magnetization is rotating in plane, which means that $\phi = 0$ and $\phi = 90^\circ$ denotes +$x$ and +$y$ magnetization, respectively. The spin accumulation is calculated at $E_{F}$ = -0.4 eV under a reference electric field 1 V/\AA along +$x$ direction. {Note that the magnitude of the electric field (1 V/\AA) is difficult to realize experimentally and is used only as a reference value.}  } 
\end{figure*}
\section{Edelstein effect in AB and BA-stacked bilayer 2H-${\mathrm{VTe}_{2}}$}

The above analysis of monolayer 2H-VTe$_{2}$ demonstrates that the behavior of the Edelstein effect is strongly governed by the symmetry of the system. By further engineering the crystal symmetry, additional components of the Edelstein effect may emerge, some of which can even be reversibly controlled, offering significant potential for device applications\cite{PhysRevB.106.165420,REE3}. An effective strategy to achieve this is to construct bilayer 2H-VTe$_{2}$, where the dependence of crystal symmetry—and its constraints on the Edelstein effect—on the stacking configuration warrants systematic investigation. Here, we consider two stacking configurations of bilayer 2H-VTe$_{2}$: AB stacking, in which the V atoms in the top layer align with the upper Te atoms in the bottom layer [Fig. \ref{4}(a)], and BA stacking [Fig. \ref{4}(b)], where the V atoms in the bottom layer align with the lower Te atoms in the top layer. Previous studies have shown that these two configurations, which are related by an $M_z$ operation, both correspond to local energy minima and can be reversibly switched via sliding ferroelectricity\cite{VStrain}. It has also been reported that the interlayer magnetic coupling in both AB and BA stackings lies close to the ferromagnetic–antiferromagnetic (FM-AFM) phase boundary  and can be tuned toward a FM state via slight strain\cite{VStrain}. In this work, we focus on the Edelstein effect in the interlayer ferromagnetic configuration of bilayer 2H-VTe$_{2}$, in order to isolate the role of stacking-dependent crystal symmetry under a fixed magnetic background. The MAE of bilayer 2H-VTe$_{2}$ is calculated to be $E_{x}-E_{z}=E_{y}-E_{z}$=-1.88meV per V atom, indicating an easy-plane magnetization. The electronic band structure of bilayer 2H-VTe$_{2}$ is shown in Fig. \ref{4}(c). Compared with the monolayer case, the band structure becomes more complex due to the coexistence of electronic states from the two VTe$_{2}$ layers. Nevertheless, similar to the monolayer, the valence band maximum (VBM) is located at the $\Gamma$ point, while the conduction band minimum (CBM) resides at the $K$ point, resulting in a small indirect band gap of approximately 0.06 eV.

Having understood the basic electronic properties, we now work on qualitatively exploring the Edelstein effect in bilayer  2H-VTe$_{2}$ from the viewpoint of symmetry. For both AB and BA stackings of bilayer 2H-VTe$_{2}$, the crystal symmetry belongs to the $C_{3v}$ point group, which allows only the $\delta S_y^{\mathrm{even}}$ component under an applied electric field along the +$x$ direction. Upon introducing in-plane ferromagnetic order, the symmetry is further reduced, thereby enabling additional components of the current-induced spin accumulation. For example, when the magnetization is aligned along the +$x$ direction, the magnetic point group reduces to $m$, with symmetry generators $\{I,M_{x}\}$. The $M_x$ symmetry operation forbids both $\chi_{xx}^{\mathrm{even}}$and $\chi_{xx}^{\mathrm{odd}}$, while allowing other components associated with an electric field along the +$x$ direction, namely $\chi_{yx}^{\mathrm{even}}$, $\chi_{zx}^{\mathrm{even}}$, $\chi_{yx}^{\mathrm{odd}}$, and $\chi_{zx}^{\mathrm{odd}}$. Consequently, under an electric field applied along the +$x$ direction, both time-reversal-even and -odd components of $\delta S_y$ and $\delta S_z$ can emerge. In contrast, when the magnetization is aligned along the +$y$ direction, the magnetic point group becomes $m^\prime$, characterized by $\{I,M_{x}T\}$. Under the $M_xT$ symmetry, $\chi_{xx}^{\mathrm{even}}$, $\chi_{yx}^{\mathrm{odd}}$, and $\chi_{zx}^{\mathrm{odd}}$ are forbidden, while the remaining components associated with +$x$ electric field are allowed. As a result, for an electric field along the +$x$ direction, $\delta S_y^{\mathrm{even}}$, $\delta S_z^{\mathrm{even}}$, and $\delta S_x^{\mathrm{odd}}$are permitted. Notably, this symmetry-based analysis is general and can be extended to other bilayer 2H-MX$_{2}$ systems with in-plane magnetization.

To further quantify the symmetry-allowed current-induced spin components, we calculate the spin accumulation per unit electric field (V/\AA) as a function of the Fermi energy for both AB and BA stackings, as shown in Fig. \ref{5}. For the case of magnetization along the +$x$ direction, the time-reversal-even component $\delta S_z^{\mathrm{even}}$ exhibits a peak value of approximately $-0.8 \hbar$ near $E_F=-0.4$ eV. Meanwhile, a finite $\delta S_y^{\mathrm{even}}$, which is absent in monolayer 2H-VTe$_{2}$ but becomes symmetry-allowed in the bilayer, also emerges and reaches a local maximum of 0.28 $\hbar$ near the same energy, which is less than half of $\delta S_z^{\mathrm{even}}$. Microscopically, these local maxima of spin accumulation primarily originate from electronic states near the $\Gamma$ point (see Fig. \ref{9} in Appendix D), with only minor contributions arising from states near the $K$ point. The time-reversal-odd components, $\delta S_y^{\mathrm{odd}}$ and $\delta S_z^{\mathrm{odd}}$, increase in magnitude as the Fermi energy shifts downward toward the valence band. For magnetization along the +$y$ direction, both $\delta S_y^{\mathrm{even}}$ and $\delta S_z^{\mathrm{even}}$ exhibit pronounced peaks of approximately 1.0 $\hbar$ and 1.2 $\hbar$, respectively, near $E_F=-0.4$ eV, which are again originated from the electronic states near $\Gamma$ point (see Fig. \ref{9} in Appendix D). In contrast, $\delta S_x^{\mathrm{odd}}$ increases and gradually saturates at around -0.5 $\hbar$ in the same energy region. Notably, switching the stacking configuration from AB to BA—achievable via ferroelectric-controllable interlayer sliding and equivalent to applying an $M_zT$ operation under in-plane magnetization—leads to a sign reversal of $\delta S_y^{\mathrm{even}}$ and $\delta S_z^{\mathrm{odd}}$, while the other nonzero components remain unchanged. Similar to the case of monolayer VTe$_{2}$, the spin accumulation in bilayer VTe$_{2}$ remains negligible when the Fermi level is shifted upward from the band gap into the low-energy conduction-band region. This behavior arises because the electronic states near the $K$ point, which dominate the low-energy conduction bands, make negligible contributions to the net spin accumulation (see Fig. \ref{10} in Appendix D). Although such sliding-induced sign reversal of current-induced spins has been previously reported for the time-reversal-even contribution in nonmagnetic bilayer WTe$_{2}$\cite{FE4}, its realization in magnetic systems—particularly for both even and odd components—remains largely unexplored and is of great relevance for nonvolatile control of spin–orbit torque. From a symmetry perspective, this selective sign reversal of current-induced spin components originates from the fact that the $M_zT$ operation changes the sign of $\chi_{yx}^{\mathrm{even}}$ and $\chi_{zx}^{\mathrm{odd}}$ under in-plane magnetization, thereby reversing $\delta S_y^{\mathrm{even}}$ and $\delta S_z^{\mathrm{odd}}$. These results highlight the potential of manipulating charge–spin conversion in two-dimensional systems via stacking engineering.

The above analysis demonstrates that the current-induced spin accumulation is highly sensitive to the magnetization direction, motivating a detailed investigation of the magnetization-dependent Edelstein effect in bilayer 2H-VTe$_{2}$. To this end, we calculate the angular dependence of the current-induced spin accumulation for AB-stacked bilayer 2H-VTe$_{2}$ at $E_F=-0.4$ eV under per applied driven electric field (V/\AA) along the +$x$ direction, as shown in Figs. \ref{6}(a) and \ref{6}(b). For the time-reversal-even contribution, all three components of $\delta S^{\mathrm{even}}$ emerge during the rotation of the in-plane magnetization. This contrasts with the monolayer case, where only $ \delta S_z^{\mathrm{even}}$ is present. Quantitatively, both $\delta S_y^{\mathrm{even}}$and $\delta S_z^{\mathrm{even}}$ are even functions with respect to $\phi-{90}^\circ$, reaching maximum magnitudes of approximately -1.0 $\hbar$ and 1.2 $\hbar$ at $\phi={90}^\circ$, respectively. In contrast, $\delta S_x^{\mathrm{even}}$has a smaller magnitude and exhibits an odd dependence on $\phi-{90}^\circ$. For the time-reversal-odd contribution, all three components of $\delta S^{\mathrm{odd}}$ are also present, in contrast to the monolayer case where $\delta S_z^{\mathrm{odd}}$ is absent for in-plane magnetization. Among them, $\delta S_z^{\mathrm{odd}}$ has a relatively small magnitude (always below 0.2 $\hbar$), whereas $\delta S_y^{\mathrm{odd}}$ and $\delta S_z^{\mathrm{odd}}$ can reach values up to approximately 0.4 $\sim$ 0.5 $\hbar$. In terms of angular dependence, $\delta S_y^{\mathrm{odd}}$ and $\delta S_z^{\mathrm{odd}}$ are odd functions of $\phi-{90}^\circ$, while $\delta S_x^{\mathrm{odd}}$ is an even function. Overall, the magnitude of the current-induced spin accumulation is on the order of 0.1 $\sim$ 1 $\hbar$ per V/\AA. By normalizing with the calculated conductivity and assuming an effective thickness of 12 \AA for the bilayer, we estimate that a current density of ${10}^7$ A/cm$^2$ induces magnetic moments on the order of ${10}^{-7}$to ${10}^{-6}$ $\hbar$ per unit cell (see Appendix A). To understand the parity dependence of the current-induced spin components on the magnetization angle, we note that changing the magnetization angle from$ \phi$ to ${180}^\circ-\phi$ is equivalent to applying the $M_xT$ operation. Under this symmetry, $\chi_{yx}^{\mathrm{even}}$, $\chi_{zx}^{\mathrm{even}}$, and $\chi_{xx}^{\mathrm{odd}}$remain invariant, whereas $\chi_{xx}^{\mathrm{even}}$, $\chi_{yx}^{\mathrm{odd}}$, and $\chi_{zx}^{\mathrm{odd}}$change sign. Consequently, for an electric field applied along the +$x$ direction, $\delta S_y^{\mathrm{even}}$, $\delta S_z^{\mathrm{even}}$, and $\delta S_x^{\mathrm{odd}}$(proportional to $\chi_{yx}^{\mathrm{even}}$,$ \chi_{zx}^{\mathrm{even}}$, and $\chi_{xx}^{\mathrm{odd}}$) are even functions of $\phi-{90}^\circ$, whereas $\delta S_x^{\mathrm{even}}$, $\delta S_y^{\mathrm{odd}}$, and $\delta S_z^{\mathrm{odd}}$(proportional to $\chi_{xx}^{\mathrm{even}}$, $\chi_{yx}^{\mathrm{odd}}$, and $\chi_{zx}^{\mathrm{odd}}$) are odd functions. Notably, when the stacking configuration is switched from AB to BA, $\delta S_x^{\mathrm{even}}$, $\delta S_y^{\mathrm{even}}$, and $\delta S_z^{\mathrm{odd}}$reverse their signs over the entire range of magnetization angles, while the other components remain unchanged (see Appendix E). This selective sign reversal originates from the fact that switching from AB to BA under in-plane magnetization is equivalent to applying an $M_zT$ operation. Under this symmetry, $\chi_{xx}^{\mathrm{even}}$, $\chi_{yx}^{\mathrm{even}}$, and $\chi_{zx}^{\mathrm{odd}}$change sign, whereas $\chi_{xx}^{\mathrm{odd}}$, $\chi_{yx}^{\mathrm{odd}}$, and $\chi_{zx}^{\mathrm{even}}$remain invariant. Therefore, for an electric field applied along the +$x$ direction, the corresponding spin components $\delta S_x^{\mathrm{even}}$, $\delta S_y^{\mathrm{even}}$, and $\delta S_z^{\mathrm{odd}}$reverse sign, while the remaining components are unchanged.

\section{conclusions}
In conclusion, based on first-principles calculations and symmetry analysis, we systematically investigate the magnetization-dependent Edelstein effect in monolayer and bilayer ferromagnetic 2H-VTe$_{2}$. For monolayer 2H-VTe$_{2}$ with $D_{3h}$ crystal symmetry, we find that when the magnetization is aligned along the +$x$ (+$y$) direction, only $\delta S_z^{\mathrm{even}}$and $\delta S_y^{\mathrm{odd}}$($\delta S_z^{\mathrm{even}}$ and $\delta S_x^{\mathrm{odd}}$) are allowed under an applied electric field along the +$x$ direction. For bilayer 2H-VTe$_{2}$ in AB and BA stackings under an applied electric field along +$x$ direction, where the crystal symmetry is reduced to $C_{3v}$, additional components emerge. Specifically, for magnetization along +$x$ (+$y$), besides $\delta S_z^{\mathrm{even}}$and $\delta S_y^{\mathrm{odd}}$($\delta S_z^{\mathrm{even}}$ and $\delta S_x^{\mathrm{odd}}$), extra components such as $\delta S_y^{\mathrm{even}}$and $\delta S_z^{\mathrm{odd}}$($\delta S_y^{\mathrm{even}}$) become allowed, some of which can be reversibly controlled by switching between AB and BA stackings. Furthermore, upon rotating the in-plane magnetization, both monolayer and bilayer systems exhibit symmetry-constrained angular dependence of the current-induced spin accumulation. Notably, the symmetry-governed and angle-dependent qualitative behavior of the Edelstein effect identified in 2H-VTe$_{2}$ can be generalized to other 2H-MX$_{2}$ ferromagnetic systems with the same symmetry, highlighting this material family as a promising platform for efficient charge–spin conversion. Our findings not only deepen the understanding of current-induced spin accumulation in magnetic systems, but also pave the way for realizing intrinsic and tunable Edelstein effects for future spin–orbit torque devices.

\section{Ackownledgements}
This project was supported by the European Union Graphene Flagship project 2DSPIN-TECH (grant agreement No. 101135853) and SFB 1277 (Project-ID 314695032).

\section{Appendix A: Normalizing Edelstein coefficient with conductivity}

Besides the original Edelstein effect 
$\delta{S}_{i} = \chi_{ij} E_{j}$ used in the main text, which denotes the number of spin angular momentum ($\hbar$) in a single unit cell induced by per electric field (1 V/\AA) and
has the unit of $\hbar$ \AA/V, one can also define a normalized current-induced Edelstein effect, $\delta{S}_{i} = \chi'_{ij} J_{j} $, which denotes the three-dimensional volume density of spins angular momentum ($\hbar/\textup{cm}^{3}$) induced by per current density ($\textup{A}/\textup{cm}^{2}$) and possesses the unit of $\hbar/(\textup{A}\cdot \textup{cm})$.

Since the electric current component $J_{i}$ can be related with electric field $E_{i}$ with $J_{i} = \sigma_{ii}E_{i}$, where $\sigma_{ii}$ is the longitudinal conductivity of the system, the $\chi_{ij}$ can be thus converted into $\chi'_{ij}$ with the following expression:

\begin{equation}
    \chi'_{ij} = \chi_{ij} /(V \cdot \sigma_{ii} )
\end{equation}

where $V$ is the effective volume of the system. Here $\sigma_{ii}$ can be calculated using the Kubo formula implemented in Linres code:

\begin{equation}
    \sigma_{ii} = \frac{e^{2}\hbar}{\pi V N} \sum_{\mathbf{k},m,n} \frac{\Gamma^{2} \textup{Re}(\langle n\mathbf{k} |\hat{v}_{i} | m\mathbf{k} \rangle \langle m\mathbf{k} | \hat{v}_{i} | n\mathbf{k} \rangle  )  }{[(E_{f}-\epsilon_{n\mathbf{k}})^{2}+ \Gamma^{2}][(E_{f}-\epsilon_{m\mathbf{k}})^{2}+ \Gamma^{2}]}
\end{equation}

Here, $e$ is the elementary charge; $n$ and $m$ denote band indices; $\textbf{k}$ is the Bloch wave vector; $E_{F}$ is the Fermi energy; $\hat{v}$ is the velocity operator, $\hat{S}$ is the spin operator; $\epsilon_{n\textbf{k}}$ is the eigenvalue; $V$ is the volume of unit cell after considering the effective thickness of two-dimensional monolayer; $N$ is the total number of k points used to sample the Brillouin zone; and $\Gamma$ = 10 meV is the disorder parameter. Based on the calculated longitude conductivity and considering the thickness of monolayer and bilayer 2H-VTe$_{2}$ for about 6 \AA and 12 \AA, one can find that a $\chi_{ij}$ = 1$ \hbar $\AA/V in per unit cell corresponds to 8$\times10^9 \hbar$/ (A$\cdot$ cm) and 4$\times10^9 \hbar$/ (A$\cdot$ cm), respectively. By further considering the strength of driven current and the effective volume of the system, one can arrive at the amount of spin accumulation under certain current. The magnitude of the normalized time-reversal-even Edelstein effect, which dominates the total Edelstein effect in our system, is largely independent of the $\Gamma$ broadening and can thus serve as a reliable reference value.

\section{Appendix B: k-resolved Edelstein effect pattern for monolayer 2H-${\mathrm{VTe}_{2}}$}

\begin{figure*}[ht]
\includegraphics[scale = 0.42 ]{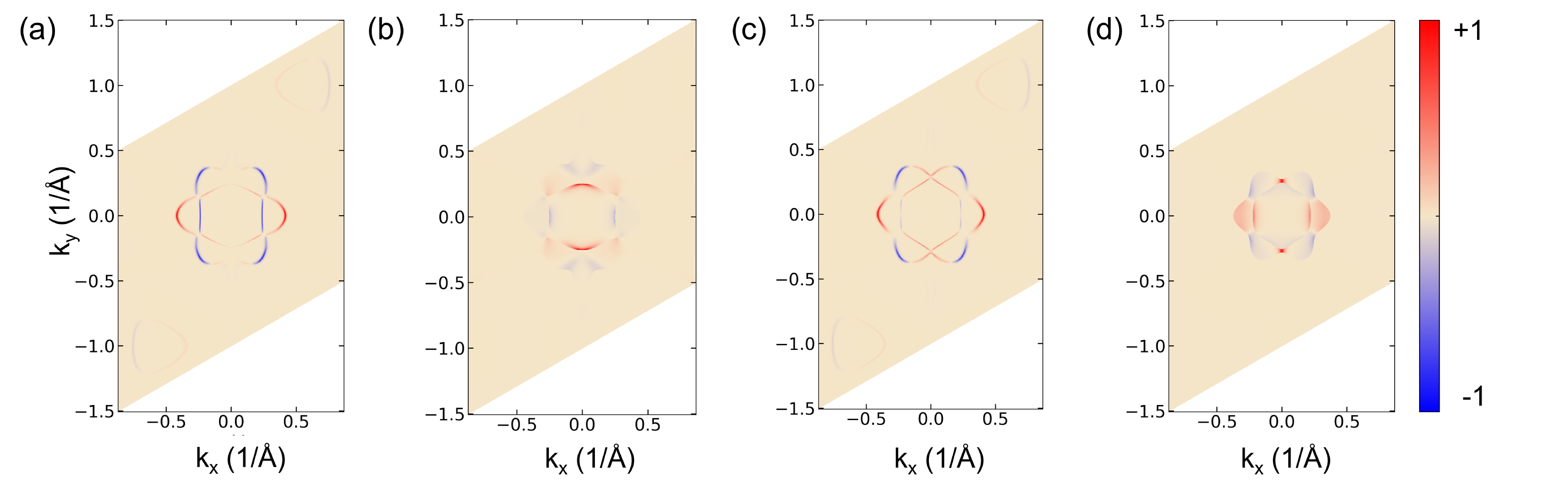}
\caption{\label{7} The k-resolved current-induced spin accumulation under a fixed reference electric field of 1V/\AA along +$x$ direction for monolayer 2H-VTe$_{2}$. (a) denotes $\delta S^{\textup{even}}_{z}$ at $E_{\textup{F}}$ = -0.44 eV for magnetization along +$x$,  (b) denotes $\delta S^{\textup{odd}}_{z}$ at $E_{\textup{F}}$ = -0.47 eV for magnetization along +$x$, (c) denotes $\delta S^{\textup{even}}_{z}$ at $E_{\textup{F}}$ = -0.44 eV for magnetization along +$y$, (d) denotes $\delta S^{\textup{odd}}_{z}$ at $E_{\textup{F}}$ = -0.4 eV for magnetization along +$y$. All selected $E_{\textup{F}}$ values are located at local maxima of the calculated spin accumulation. The red and blue denotes the positive and negative values of spin accumulations, respectively, which have been normalized to the maxima value.   } 
\end{figure*}

\begin{figure*}[ht]
\includegraphics[scale = 0.3 ]{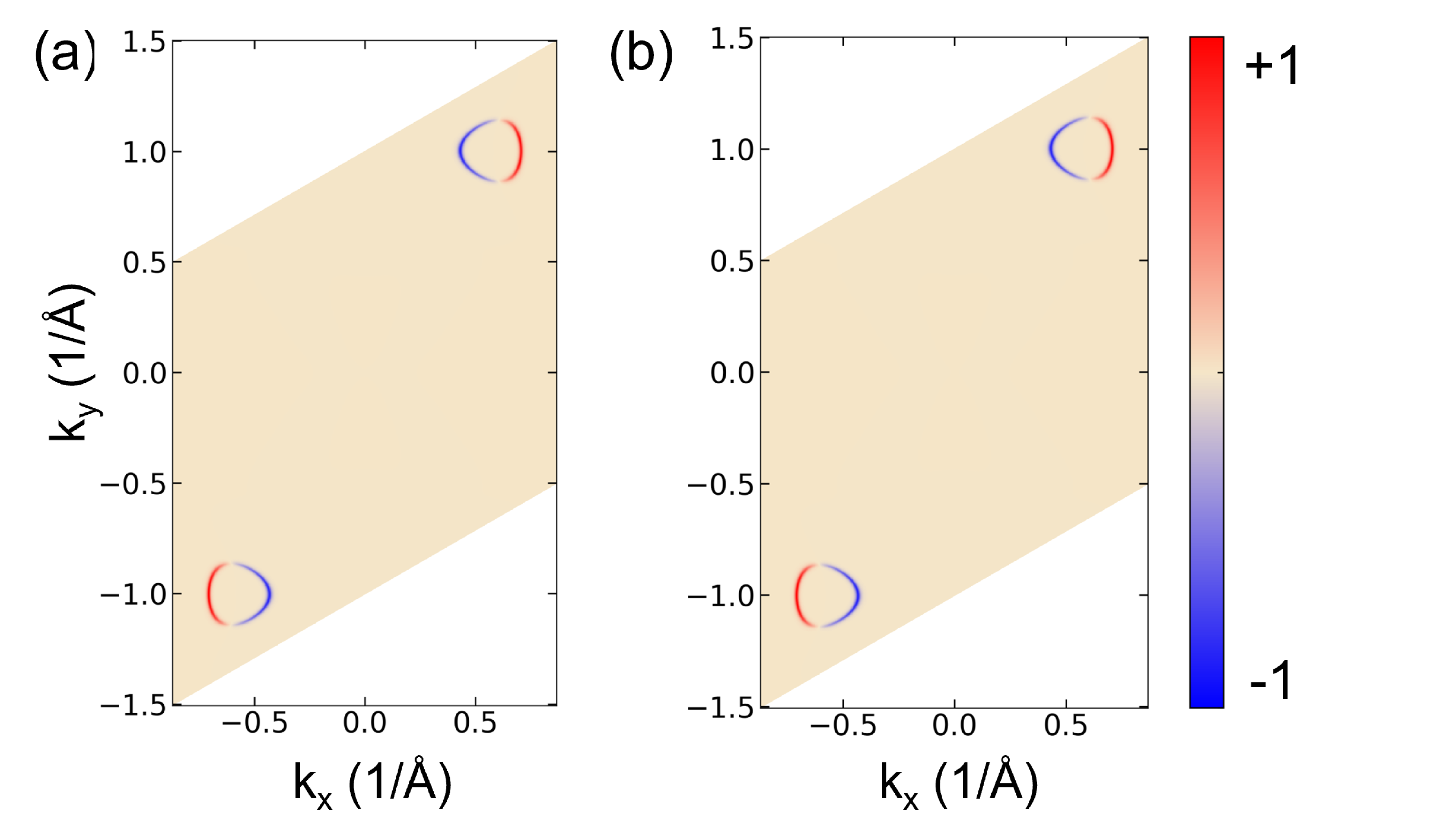}
\caption{\label{8} The k-resolved current-induced spin accumulation under a fixed reference electric field of 1V/\AA along +$x$ direction for monolayer 2H-VTe$_{2}$ under $E_{\textup{F}}$ = 0.2 eV. (a) denotes $\delta S^{\textup{even}}_{z}$ for magnetization along +$x$,  (b) denotes $\delta S^{\textup{even}}_{z}$ at for magnetization along +$y$. The red and blue denotes the positive and negative values of spin accumulations, respectively, which have been normalized to the maxima value.   } 
\end{figure*}

In Fig. \ref{7} we show the  k-resolved current-induced spin accumulation components under a fixed +$x$ electric field ( 1 V/\AA) for monolayer 2H-VTe$_{2}$ at Fermi levels corresponding to the local maxima in the low-energy valance band region. In Fig.\ref{8} we show the k-resolved current-induced spin accumulation components at $E_{\textup{F}}$ = 0.2 eV, corresponding to the low-energy conduction band region. Note that for $E_{\textup{F}}$ = 0.2 eV, the time-reversal-odd spin accumulation components vanish throughout the entire $k$-space and are therefore not plotted.

\section{Appendix C: Angular-dependence of $\delta S^{\textup{even}}_{z}$ for monolayer 2H-${\mathrm{VTe}_{2}}$}
To qualitatively understand why $\delta S^{\textup{even}}_{z}$ is roughly proportional to cos$2\phi$, one should figure out the expression of $\delta S^{\textup{even}}_{z}$ as a function of in-plane magnetization direction in the second-order dependence, which is generally expressed as:
\begin{equation}
    \delta S^{\textup{even}}_{z} = \sum_{\alpha,\beta} c_{\alpha \beta}m_{\alpha}m_{\beta} E_{x} + \sum_{\alpha,\beta} c'_{\alpha \beta}m_{\alpha}m_{\beta} E_{y}
\end{equation}
where $\alpha$ and $\beta$ denotes the in-plane coordinate $x$ or $y$, and $m$ = ($m_{x}$, $m_{y}$) denotes the unit vector along the in-plane magnetization direction. 
By applying the symmetry operations of monolayer 2H-VTe$_{2}$ ($C_{3}$, $M_{z}$ and $M_{x}$ in  $D_{3h}$ point group symmetry) on the above equation, the expression will still be satisfied, which allow us to rule out the symmetry-forbidden terms and thereby lead to a clearer expression. Such approach has been adopted to arrive at the low-order magnetization dependence behavior of Edelstein effect in magnetic systems \cite{FGT3,MX2_1}.

The above mentioned symmetry operations will transform the $E_{x}$, $m_{x}$, $m_{y}$ and $\delta S^{\textup{even}}_{z}$ as:
\begin{equation}
\begin{aligned}
C_{3}: \quad
&E_{x} \longrightarrow -\frac{1}{2}E_{x}
-\frac{\sqrt{3}}{2}E_{y},
\qquad
E_{y} \longrightarrow \frac{\sqrt{3}}{2}E_{x}
-\frac{1}{2}E_{y}, \\
&m_{x} \longrightarrow -\frac{1}{2}m_{x}
-\frac{\sqrt{3}}{2}m_{y},
\qquad
m_{y} \longrightarrow \frac{\sqrt{3}}{2}m_{x}
-\frac{1}{2}m_{y}, \\
&\delta S^{\mathrm{even}}_{z}
\longrightarrow
\delta S^{\mathrm{even}}_{z}, \\[6pt]
M_{x}: \quad
&E_{x} \longrightarrow -E_{x},
\qquad
E_{y} \longrightarrow E_{y}, \\
&m_{x} \longrightarrow m_{x},
\qquad
m_{y} \longrightarrow -m_{y}, \\
&\delta S^{\mathrm{even}}_{z}
\longrightarrow
-\delta S^{\mathrm{even}}_{z}, \\[6pt]
M_{z}: \quad
&E_{x} \longrightarrow E_{x},
\qquad
E_{y} \longrightarrow E_{y}, \\
&m_{x} \longrightarrow -m_{x},
\qquad
m_{y} \longrightarrow -m_{y}, \\
&\delta S^{\mathrm{even}}_{z}
\longrightarrow
\delta S^{\mathrm{even}}_{z}.
\end{aligned}
\end{equation}

By considering the constraint of $M_{x}$ and $M_{z}$, one can arrive at: 
\begin{equation}
\delta s_z^{\textup{even}} =
\left(
c_{xx}m_x^2 + c_{yy}m_y^2
\right)E_x
+ c'_{xy}m_xm_yE_y .
\end{equation}

Further considering the invariance under $C_{3}$ would lead to: 
\begin{equation}
c_{xx} = -c_{yy} = -\frac{c'_{xy}}{2},
\qquad
c_{zz} = 0.
\end{equation}

Therefore, the final expression of $\delta S^{\textup{even}}_{z}$ is:
\begin{equation}
\delta S_z^{\textup{even}}
=
c\left[
\left(m_x^2-m_y^2\right)E_x
-2m_xm_yE_y
\right] .
\end{equation}

Based on the above expression, one can realize that when electric field is along $+x$ direction, the $\delta S^{\textup{even}}_{z}$ under in-plane magnetization along ($m_{x}$, $m_{y}$) = (cos$\phi$, sin$\phi$) is propotional to cos2$\phi$, which is very close to the behavior shown in Fig. \ref{3}(a) in the main text. This means the magnetization dependence of $\delta S^{\textup{even}}_{z}$ in monolayer 2H-VTe$_{2}$ is dominated by the second order contribution, while the contributions from the higher-order terms are very small.

\section{Appendix D: k-resolved Edelstein effect pattern for AB-stacked bilayer 2H-${\mathrm{VTe}_{2}}$}

\begin{figure*}[ht]
\includegraphics[scale = 0.42 ]{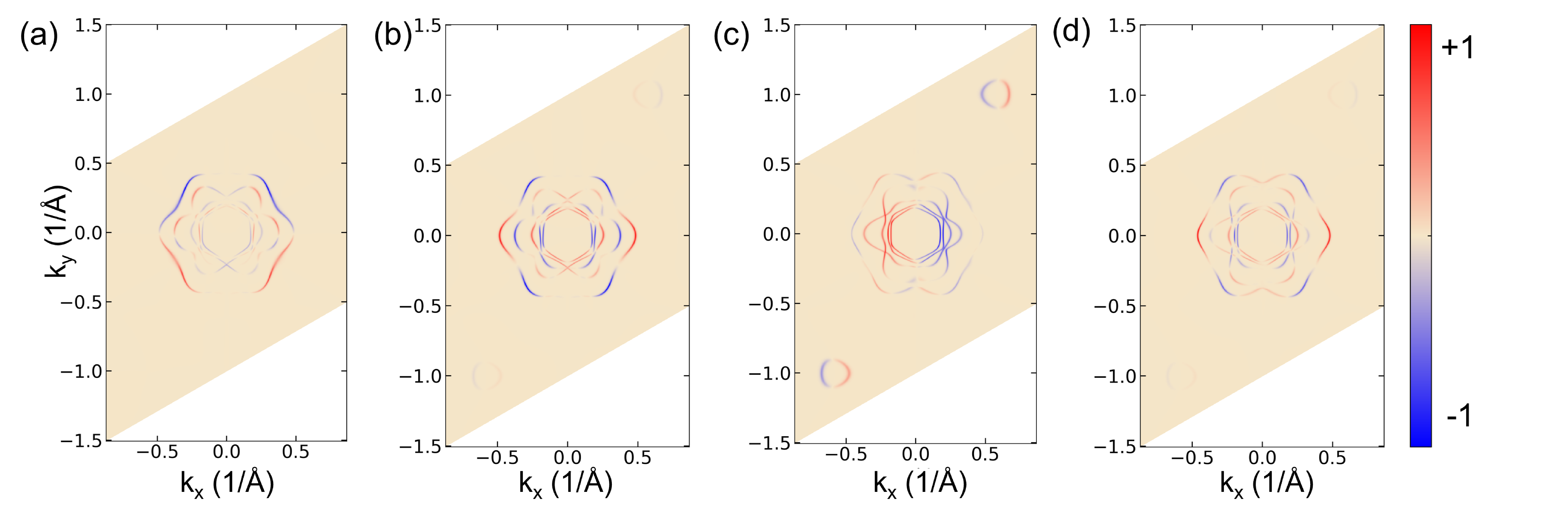}
\caption{\label{9} The k-resolved current-induced spin accumulation under a fixed reference electric field of 1V/\AA along +$x$ direction for bilayer 2H-VTe$_{2}$. (a) denotes $\delta S^{\textup{even}}_{z}$ at $E_{\textup{F}}$ = -0.4 eV for magnetization along +$x$,  (b) denotes $\delta S^{\textup{odd}}_{z}$ at $E_{\textup{F}}$ = -0.4 eV for magnetization along +$x$, (c) denotes $\delta S^{\textup{even}}_{z}$ at $E_{\textup{F}}$ = -0.4 eV for magnetization along +$y$, (d) denotes $\delta S^{\textup{odd}}_{z}$ at $E_{\textup{F}}$ = -0.4 eV for magnetization along +$y$. All selected $E_{\textup{F}}$ values are located at local maxima of the calculated spin accumulation. The red and blue denotes the positive and negative values of spin accumulations, respectively, which have been normalized to the maxima value.   } 
\end{figure*}

\begin{figure*}[ht]
\includegraphics[scale = 0.42 ]{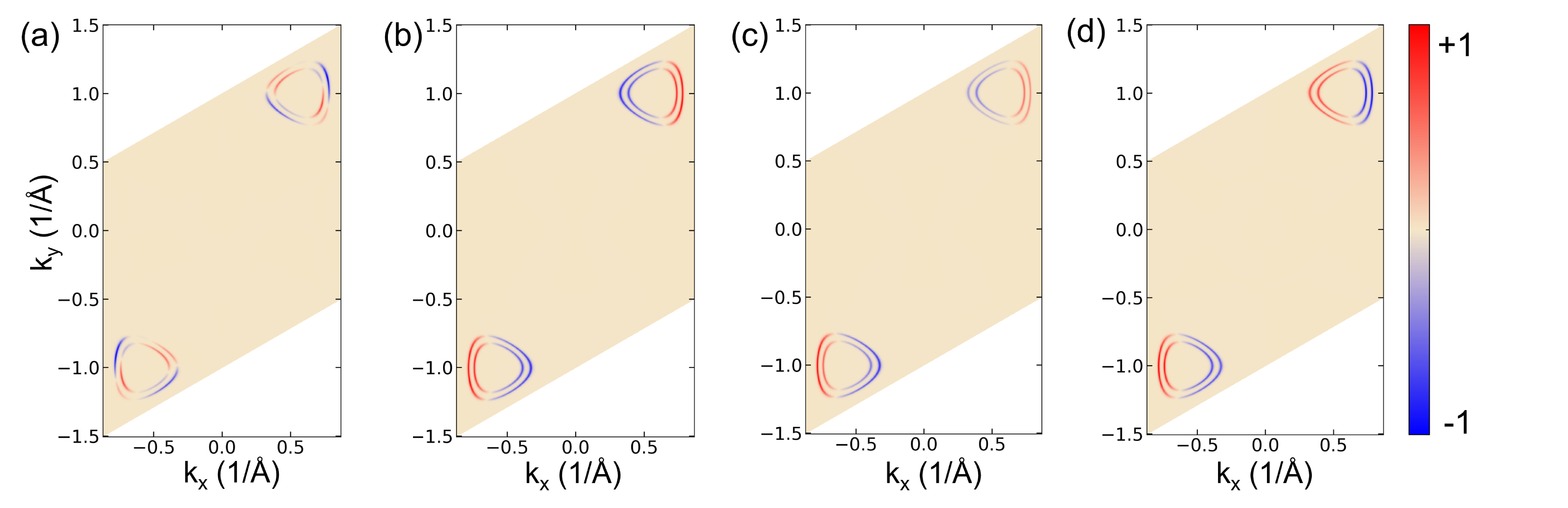}
\caption{\label{10} The k-resolved current-induced spin accumulation under a fixed reference electric field of 1V/\AA along +$x$ direction for AB-stacked bilayer 2H-VTe$_{2}$ under $E_{\textup{F}}$ = 0.2 eV. (a) denotes $\delta S^{\textup{even}}_{z}$ for magnetization along +$x$,  (b) denotes $\delta S^{\textup{odd}}_{z}$ for magnetization along +$x$, (c) denotes $\delta S^{\textup{even}}_{z}$ for magnetization along +$y$, (d) denotes $\delta S^{\textup{odd}}_{z}$ for magnetization along +$y$. The red and blue denotes the positive and negative values of spin accumulations, respectively, which have been normalized to the maxima value. } 
\end{figure*}

In Fig. \ref{9} we show the  k-resolved current-induced spin accumulation components under a fixed +$x$ electric field ( 1 V/\AA) for AB-stacked bilayer 2H-VTe$_{2}$ at Fermi levels corresponding to the local maxima in the low-energy valance band region. In Fig.\ref{10} we show the k-resolved current-induced spin accumulation components at $E_{\textup{F}}$ = 0.2 eV, corresponding to the low-energy conduction band region. Note that for $E_{\textup{F}}$ = 0.2 eV, the time-reversal-odd spin accumulation components vanish throughout the entire $k$-space and are therefore not plotted.

\section{Appendix E: Angular-dependent current-induced spin accumulation for bilayer 2H-${\mathrm{VTe}_{2}}$ with BA stacking}
\begin{figure*}[ht]
\includegraphics[scale = 0.42 ]{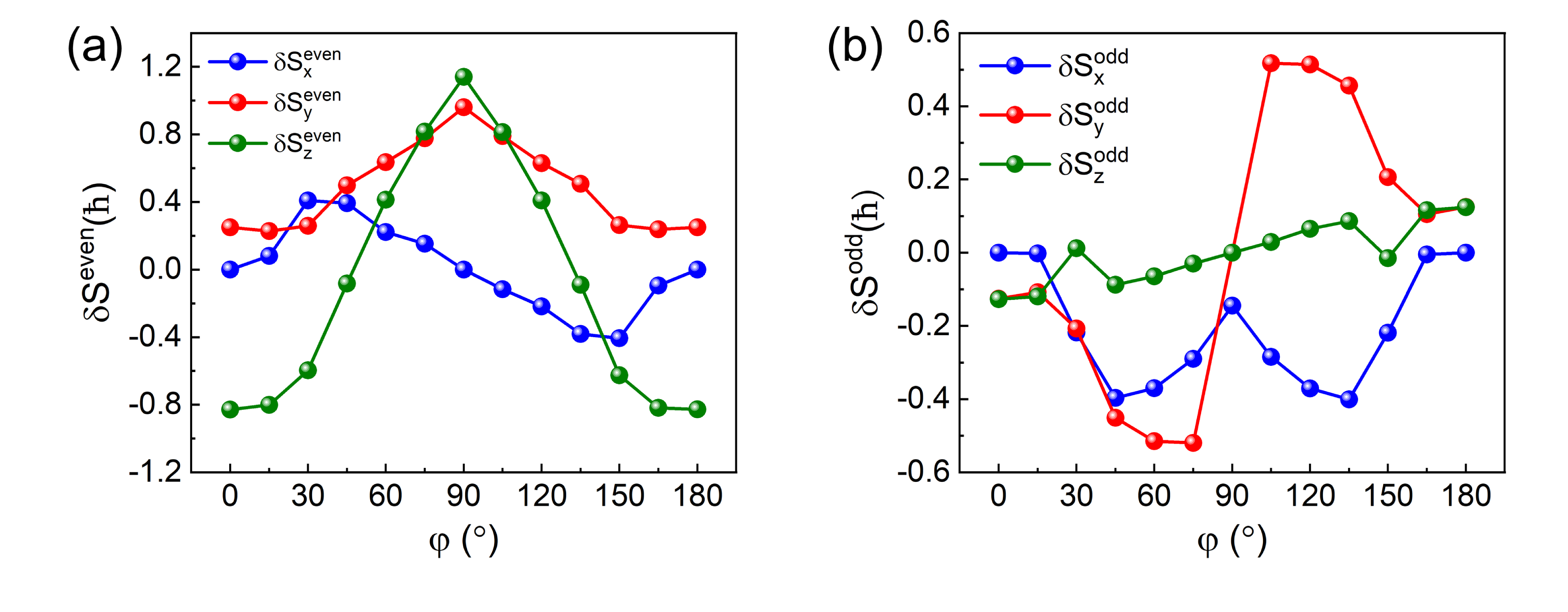}
\caption{\label{11} Angular-dependent spin accumulation of BA-stacked bilayer 2H-VTe$_{2}$, which is denoted as a function of $\psi$ (angle between magnetization direction and +$x$ direction). (a) denotes the symmetry-allowed time-reversal-even components (only $\delta S_z^{\mathrm{even}}$ here) of spin accumulation in a unit cell as function of $\psi$, while (b) denotes the symmetry-allowed time-reversal-odd components of spin accumulation in a unit cell as function of $\psi$. Note that the magnetization is rotating in plane, which means that $\psi = 0$ and $\psi = 90^\circ$ denotes +$x$ and +$y$ magnetization, respectively. The spin accumulation is calculated under $E_{\textup{F}}$ = -0.4 eV under a reference electric field of 1 V/\AA along +$x$ direction.   }
\end{figure*}

Below we show the angular-dependent current-induced spin accumulation for bilayer 2H-VTe$_{2}$ with BA stacking, which can be seen in Fig.\ref{11}. 

\clearpage

%\bibliography{achemso-demo}
\end{document}